\documentclass[prb,twocolumn, superscriptaddress, showpacs,amsmath,amssymb,letterpaper]{revtex4}
\usepackage{graphicx}
\usepackage{dcolumn}
\usepackage{bm}

\begin{document}


\title{Anisotropy of Magnetic Interactions  in the Spin-Ladder Compound  (C$_5$H$_{12}$N)$_2$CuBr$_4$}

\author{E. \v{C}i\v{z}m\'{a}r}
\affiliation{Dresden High Magnetic Field Laboratory (HLD), FZ Dresden-Rossendorf, D-01314 Dresden, Germany }
\affiliation{Centre of Low Temperature Physics, P.J. \v{S}af\'{a}rik University, SK-041 54 Ko\v{s}ice, Slovakia}
\author{M. Ozerov}
\author{J. Wosnitza}
\affiliation{Dresden High Magnetic Field Laboratory (HLD), FZ Dresden-Rossendorf, D-01314 Dresden, Germany }
\author{B. Thielemann}
\affiliation{Laboratory for Neutron Scattering, ETH Z\"{u}rich and Paul Scherrer Institute, CH-5232 Villigen, Switzerland}
\author{K. W. Kr\"{a}mer}
\affiliation{Department of Chemistry and Biochemistry, University of Bern, CH-3000 Bern 9, Switzerland}
\author{Ch. R{\"{u}}egg}
\affiliation{London Centre for Nanotechnology and Department of Physics and Astronomy, University College London, London WC1E 6BT, United Kingdom}
\author{O. Piovesana}
\affiliation{Dipartimento di Chimica, Universit\'{a} di Perugia, I-06100 Perugia, Italy}
\author{M. Klanj{\v{s}}ek}
\affiliation{Laboratoire National des Champs Magn\'{e}tique Intenses, CNRS, BP 166, 38042 Grenoble, France}
\affiliation{Jo\v{z}ef Stefan Institute, Jamova 39, SI-1000 Ljubljana, Slovenia}
\author{M. Horvati{\'{c}}}
\author{C. Berthier}
\affiliation{Laboratoire National des Champs Magn\'{e}tique Intenses, CNRS, BP 166, 38042 Grenoble, France}
\author{S. A. Zvyagin}
\affiliation{Dresden High Magnetic Field Laboratory (HLD), FZ Dresden-Rossendorf, D-01314 Dresden, Germany }

\date{\today}

\begin{abstract} Magnetic excitations  in the  spin-ladder material (C$_5$H$_{12}$N)$_2$CuBr$_4$ [BPCB] are probed by high-resolution multi-frequency  electron spin resonance (ESR) spectroscopy. Our experiments provide a~direct evidence for a~biaxial anisotropy ($\sim 5\%$ of the dominant exchange interaction), that is in contrast to a fully isotropic spin-ladder model employed for this system previously. It is argued that this anisotropy in BPCB is caused by spin-orbit coupling, which appears to  be important for describing  magnetic properties of this compound. The zero-field zone-center gap in the excitation spectrum of BPCB, $\Delta_0/k_{B}=16.5$~K, is detected directly. Furthermore, an ESR signature of the inter-ladder exchange interactions  is obtained. The detailed characterization of the anisotropy in BPCB completes the determination of the full spin hamiltonian of this exceptional spin-ladder material and shows ways to study anisotropy effects in spin ladders.
\end{abstract}

\maketitle
\emph{Introduction.---}
Quantum spins on ladder-like structures have stimulated intense interest in low-dimensional magnetism. Apart from the possible relevance to high-temperature superconductivity~\cite{Siegrist,scond, Uehara}, the interest in spin ladders was motivated by their rich temperature-magnetic field phase diagram affected by quantum critical fluctuations.

Ideally, the magnetic properties of spin-ladder compounds~\cite{Azuma, Landee, Masuda, WatsonBPCB,NMR_BPCB,neutrondiffr,Sologubenko,BeniBPCB,Ruegg-Thermo} are described using the standard Heisenberg hamiltonian, including only isotropic exchange interactions. Nevertheless, anisotropy is $always$ present in real materials at some energy scale, and it can often be observed, for instance, in the $g$-factor. The two-leg spin-ladder Hamiltonian can then be written as
\begin{equation}
{\cal H}=J_{\|}\sum_{i,j=1,2}\textbf{S}_{i,j}\textbf{S}_{i+1,j}+J_{\bot}\sum_{i}\textbf{S}_{i,1}\textbf{S}_{i,2}+{\cal H}_{\delta},
\label{ladderham}
\end{equation}
where $J_{\bot}$ and $J_{\|}$ are isotropic exchange interactions along the ladder rungs and legs, respectively, and the term ${\cal H}_{\delta}$ represents anisotropic contributions. Theoretical and experimental studies of the anisotropy effects in spin ladders (and other quantum magnets) appear to be important topics in quantum magnetism because the anisotropy can significantly modify the ground-state properties and low-energy excitation spectrum of those systems~\cite{anisotropy1, anisotropy2, anisotropy3, anisotropy4, anisotropy5, anisotropy6, anisotropy7}.

Piperidinium copper bromide, (C$_5$H$_{12}$N)$_2$CuBr$_4$ [abbreviated as BPCB or (Hpip)$_2$CuBr$_4$], is known as a~prototypical realization of the two-leg spin-$\frac{1}{2}$ antiferromagnetic ladder system in the strong-coupling limit ($J_{\bot} > J_{\|}$)~\cite{WatsonBPCB} with an optimal energy scale for experimental investigations.
In zero magnetic field the ground state of BPCB is gapped and quantum disordered. The gap closes at $B_{c1}\approx 6.8$~T. At $B_{c2}\approx 13.8$~T, a~fully spin-polarized phase is induced.
The field-induced transition from the  quantum-disordered phase into the Luttinger-Liquid (LL) phase~\cite{LL} has been observed between $B_{c1}$ and $B_{c2}$ \cite{NMR_BPCB,neutrondiffr,Sologubenko,Ruegg-Thermo},  for the first time giving  the comprehensive insight into the Luttinger-Liquid paradigm in spin ladders. The transition into the field-induced magnetically-ordered phase at lower temperatures~\cite{NMR_BPCB,neutrondiffr} can effectively be described by employing  the Bose-Einstein condensation (BEC) formalism~\cite{BEC}. Because both critical fields, $B_{c1}$ and $B_{c2}$, can be reached using conventional superconducting magnets, BPCB offers a~unique opportunity to investigate the field-controlled evolution of the ground-state properties in spin ladders across different regions of the phase diagram.

Here, we report on a~comprehensive electron spin resonance (ESR) study of BPCB, in which we observe the distinct effect of anisotropy on the spectrum of magnetic excitations in this material. We determined the anisotropy parameters, with their total contribution amounting to approximately 0.6~K. Our findings  illuminate the importance of the anisotropy when describing the rich phase diagram of BPCB and, on the other hand, make this compound an ideal playground for studying anisotropy effects in quantum spin ladders.

\begin{figure}[t]
\includegraphics[width=0.40\textwidth,clip=true]{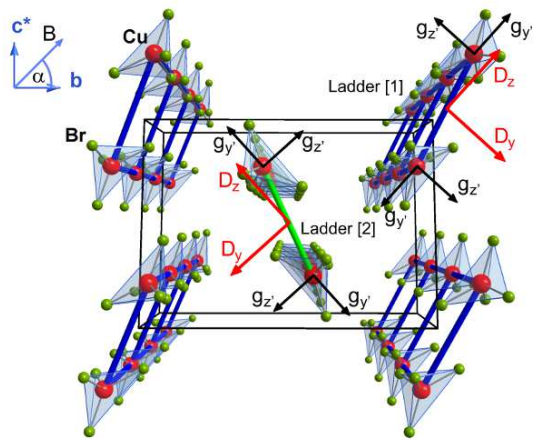}
\caption{\label{fig:bpcb} (Color online) Schematic view of the crystal structure of BPCB along the $a$~axis~\cite{Patyal}. The ladders [1] and [2] are highlighted by thick blue and green lines, respectively. Black arrows define the directions of the principal axes of the $g$-tensors, while the red  arrows define the vectors of the effective anisotropy (see text for details). Piperidinium groups are omitted for clarity.}
\end{figure}

\emph{Experiment and discussion.---}
 BPCB crystallizes in a~monoclinic lattice (space group P2$_1$/c, number of formula units per unit cell $Z=4$) with spin-$\frac{1}{2}$ Cu$^{2+}$ ions arranged in a~ladder-like structure~\cite{Patyal}. Each unit cell contains two rungs of two crystallographically equivalent ladders running along the $a$ axis (Fig.\ \ref{fig:bpcb}), related by the $c$-glide operation. Because of different orientations of principal axes of the $g$-tensor, the two sets of ladders become magnetically inequivalent in an applied magnetic field.  Exchange couplings along the rungs and legs of the ladder were determined as $J_\bot /k_B \approx~12.7 - 13.3$~K and $J_\| /k_B \approx~3.3 - 3.8$~K, respectively, depending on the experimental conditions and applied technique~\cite{WatsonBPCB,NMR_BPCB,neutrondiffr,BeniBPCB,SaviciINS,AnfusoBPCB,Lorenz,Lemmens}. A signature of a~weak inter-ladder exchange interaction ($J'\approx 100$~mK) was reported, leading to a~transition into a~field-induced magnetically-ordered phase at temperatures below 110~mK~\cite{NMR_BPCB,neutrondiffr}.

ESR experiments were performed at the Dresden High Magnetic Field Laboratory (Hochfeld-Magnetlabor Dresden) employing  an X-band spectrometer (Bruker ELEXSYS E500) at a~fixed frequency of 9.4~GHz and a~tunable-frequency ESR spectrometer (similar to that described in Ref.~\onlinecite{spectrometer}). High-quality single crystals of (C$_5$H$_{12}$N)$_2$CuBr$_4$ and its deuterated analog (C$_5$D$_{12}$N)$_2$CuBr$_4$, were used in our experiments.

\begin{figure}[t]
\includegraphics[width=0.45\textwidth,clip=true]{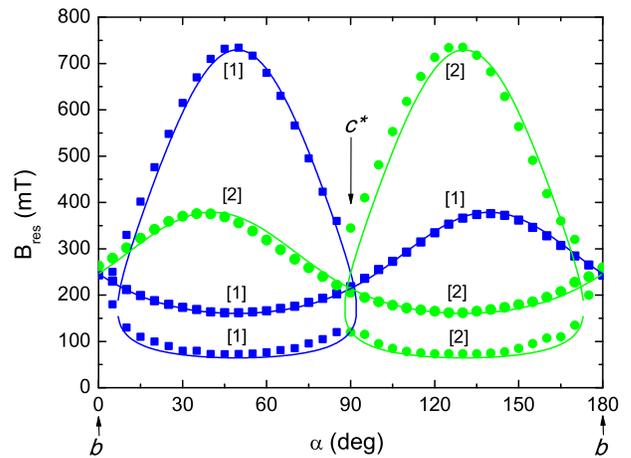}
\caption{\label{fig:XEPR} (Color online) Angular dependence of the ESR resonance fields (symbols) measured at $T=3.3$~K at 9.4~GHz, with the magnetic field applied in the $bc^{*}$ plane. Symbols labeled [1] and [2] correspond to excitations originating from the ladders [1] and [2], respectively (Fig.\ \ref{fig:bpcb}), while the lines represent the results of the simulation described in the text.}
\end{figure}

We measured the angular dependence of the ESR signal at 9.4~GHz at room temperature. We found good agreement with the results reported by Patyal \textit{et al.}~\cite{Patyal}, revealing the existence of two types of excitation centers with different principal $g$-tensor axes. However, we observed a~much more complex excitation spectrum consisting of four ESR lines at low temperatures ($T\leq7$~K)~\cite{noteD}. In Fig.\ \ref{fig:XEPR} we present the angular dependence of the resonance magnetic fields measured at 3.3~K, with the magnetic field applied in the $bc^*$ plane ($c^* \bot\:a,b$). The observed four ESR modes are incompatible with the simple isotropic spin-ladder model, where only a~single pair of modes is expected. The observation of two pairs of ESR modes  with a~pronounced angular dependence of resonance positions is a~clear signature of the presence of additional (anisotropic) interactions in BPCB. The measured temperature dependence of the integrated intensities (not presented here) indicates that these excitations correspond to the transitions within the thermally populated excited triplets. This allows us to apply a~simplified spin-triplet model with axial and in-plane anisotropy terms, $D'$ and $E'$, respectively~\cite{Bleaney}, to describe the data. The corresponding Hamiltonian for the triplet states in a magnetic field is given by
\begin{equation}
   {\cal H}_{eff}=D_xS_x^2+D_yS_y^2+D_zS_z^2-\mu_B\textbf{B}\hat{g}\textbf{S},
\label{effham}
\end{equation}
where $\mu_{B}$ is the Bohr magneton, $\textbf{B}$ is the magnetic field and $\textbf{S}$ is the triplet spin operator. The effective anisotropy parameters are defined as $D'=\frac{3}{2}D_z$ and $E'=\frac{1}{2}(D_x-D_y)$ \cite{Abragam}. We analyzed the ESR angular dependence by employing the Hamiltonian above in the ``EasySpin" simulation package~\cite{easyspin}.  The results of the simulation are presented in Fig.\ \ref{fig:XEPR} by solid lines. Good agreement with the experimental data was found for $g_{x'}=2.065$, $g_{y'}=2.045$, $g_{z'}=2.29$, $D'/k_B=0.55$~K and $E'/k_B=-0.05$~K. The anisotropy axis $D_z$ is tilted counterclockwise by $49.5^{\circ}$ from the $b$ axis in the $bc^{*}$ plane for ladder [1], and clockwise with the same angle for ladder [2] as shown in Fig.\ \ref{fig:bpcb}. The observed anisotropy energy is $\sim 5\%$ of the rung interaction $J_\bot$ (which is the dominant interaction in this compound) or $\sim 16\%$ of the leg interaction $J_\|$.

\begin{figure*}[t]
\includegraphics[width=0.65\textwidth,clip=true]{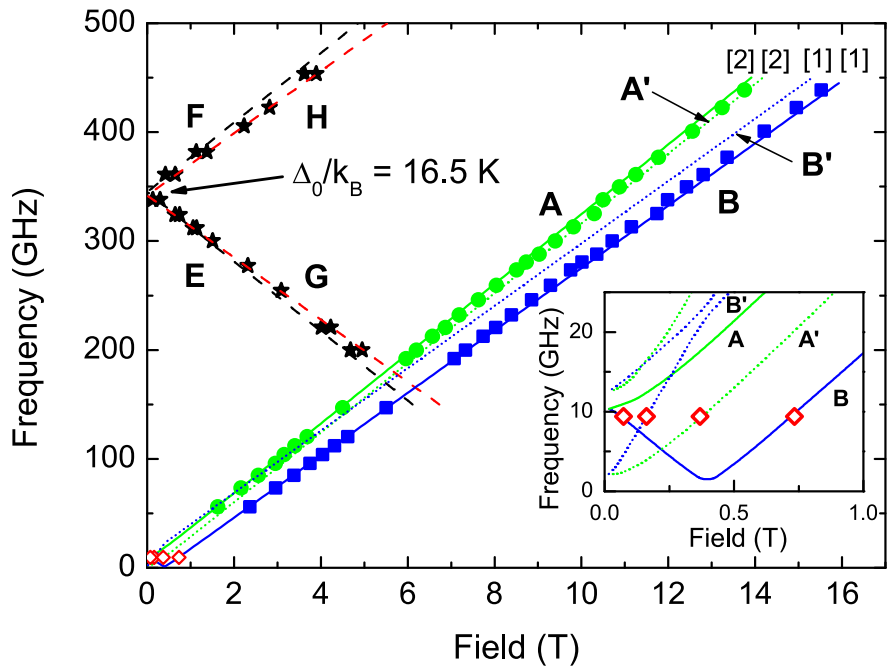}
\caption{\label{fig:EPR} (Color online) Frequency-field diagram of the ESR excitations measured at $T=1.3$ K with the magnetic field tilted by $45^{\circ}$ away from the $b$ axis in the $bc^{*}$ plane ($\alpha=45^{\circ}$). The modes \textbf{E}, \textbf{F}, \textbf{G}, and \textbf{H} (dashed lines) correspond to transitions from the spin-singlet ground state to the first excited triplet states at $k=0$. Lines [1] and [2] are results of calculations for transitions between the excited triplet levels corresponding to the ladders [1] and [2], respectively, using the set of parameters described in the text. The inset shows the low-frequency-field part of the calculated excitation diagram and experimental data obtained at 9.4~GHz and $T=3.3$~K (shown as diamonds).}
\end{figure*}

The frequency-field diagram of the magnetic excitations in BPCB was studied with the magnetic field  tilted by about $45^{\circ}$ away from the $b$ axis in the $bc^{*}$ plane (i.e., $\alpha\approx45^{\circ}$), where the best resolution is achieved (Fig.\ \ref{fig:XEPR}). Results of the experiment are presented in Fig.\ \ref{fig:EPR}.  Several important observations were made.

\begin{figure}[t]
\includegraphics[width=0.45\textwidth,clip=true]{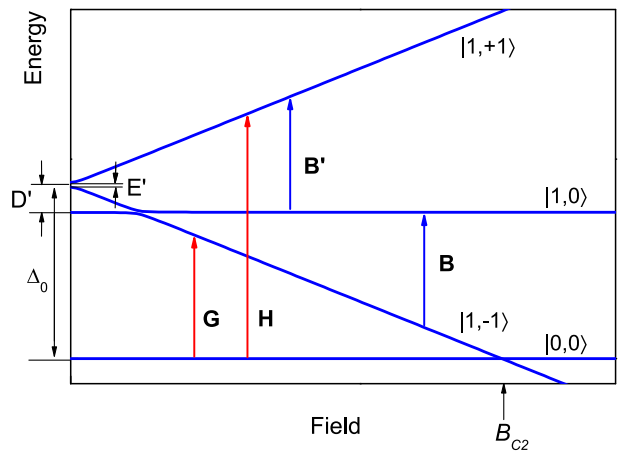}
\caption{\label{fig:levels} (Color online) Schematic energy-field diagram of ESR transitions in BPCB for ladder [1]. Excitations  from the ground state are represented by red lines (modes \textbf{G} and \textbf{H}), while transitions within excited triplets (modes \textbf{B} and \textbf{B$^{\prime}$}) are shown in green. The field orientation is the same as in Fig.\ \ref{fig:EPR}.}
\end{figure}

First, we were able to observe the spin gap in the excitation spectrum directly. The frequency-field dependency of the modes \textbf{E}, \textbf{F}, \textbf{G}, and \textbf{H} can be described by $h\nu=\Delta_0{\pm}g\mu_{B}B$, where $\Delta_0$ is the energy gap between the spin-singlet ground state and the excited spin-triplet states (Fig.\ \ref{fig:levels}). We emphasize that such ESR transitions are forbidden by selection rules if the axial symmetry is preserved \cite{Abragam} (which is clearly not the case for the chosen magnetic field orientation). The gap at zero field is 343~GHz ($\Delta_0/k_{B}=16.5$~K), which agrees well  with the value measured at the center of the Brillouin zone ($k=0$) by inelastic neutron scattering~\cite{BeniBPCB,SaviciINS}. In most gapped spin-$\frac{1}{2}$ systems, these transitions are not observable by ESR due to an overlap of the spin-triplet excitations with the two-magnon continuum. In BPCB the onset of the two-magnon continuum (with the lowest-energy boundary estimated as 2$\Delta_{\pi}/k_{B}\sim 19.2$~K, where $\Delta_{\pi}$ is the one-magnon energy gap at $k=\pi$) is at a~higher energy than the one-magnon gap at $k=0$.  This difference prevents scattering of one-magnon excitations by the two-magnon continuum and allows the observation of the gap at $k=0$.

Second, we found that the resonances are slightly split, yielding $g=2.28$ and
$g=2.04$ for the mode pairs \textbf{E}, \textbf{F} and  \textbf{G}, \textbf{H}, respectively.  A~fit of the frequency-field dependences of the observed modes reveals a~difference in zero-field splitting of about 3~GHz ($\sim$~140~mK) between the mode pairs \textbf{E}, \textbf{F} and  \textbf{G}, \textbf{H}. This is a~direct indication for the presence of inter-ladder exchange interaction and is consistent with the results of  NMR and neutron-spectroscopy experiments~\cite{NMR_BPCB, BeniBPCB}.

\begin{figure}[t]
\includegraphics[width=0.45\textwidth,clip=true]{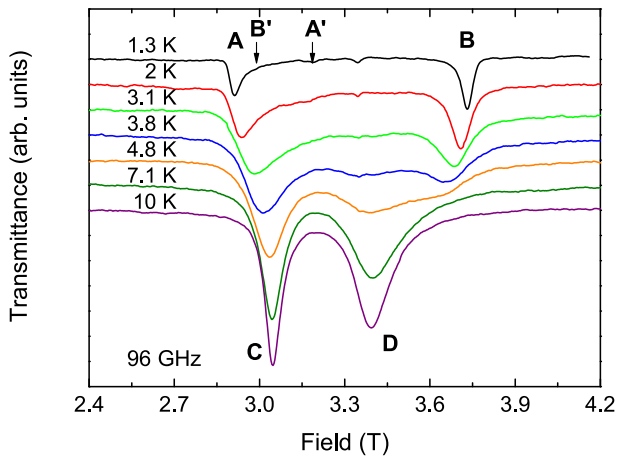}
\caption{\label{fig:spectra} (Color online) ESR spectra measured at 96~GHz with the same field orientation as in Fig.\ \ref{fig:EPR} and at temperatures as indicated. Calculated positions of the modes \textbf{A$^{\prime}$} and \textbf{B$^{\prime}$} are shown. }
\end{figure}

Third, two well-resolved  modes, \textbf{A} and \textbf{B}, with $g$-factors 2.28 and
2.04, respectively (at 1.3~K), are observed in the ESR spectra at low temperatures (Fig.\ \ref{fig:EPR}). Noticeably, the extrapolated frequency-field dependence of neither mode \textbf{A} nor mode \textbf{B} intersects zero at zero magnetic field. The calculated low-temperature frequency-field diagram, using the anisotropy parameters obtained from the analysis of the angular dependence (see Fig.\ \ref{fig:XEPR}), is presented in Fig.\ \ref{fig:EPR} by solid and dotted lines. The agreement is excellent over a~wide frequency and field range. Due to low intensity (Fig.\ \ref{fig:spectra}), modes \textbf{A$^{\prime}$} and \textbf{B$^{\prime}$} (dotted lines in Fig.\ \ref{fig:EPR}) were not observed in our high-frequency ESR experiments, which is also consistent with our calculations~\cite{intensity}. A schematic energy-field diagram of the expected ESR transitions in BPCB (for one ladder) is shown in Fig.\ \ref{fig:levels}.

\begin{figure}[t]
\includegraphics[width=0.45\textwidth,clip=true]{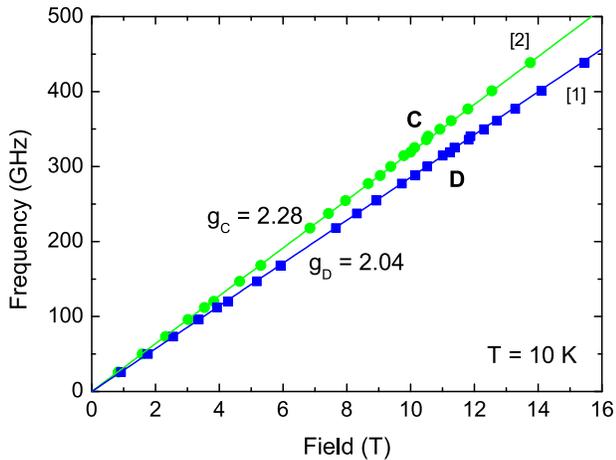}
\caption{\label{fig:EPR10} (Color online) Frequency-field diagram of the ESR excitations (symbols) measured at 10~K with the same field orientation as in Fig.\ \ref{fig:EPR}. Lines denote results of calculations
using  the formula $h\nu=g\mu_{B}B$.}
\end{figure}

The temperature evolution of the ESR absorptions measured at 96~GHz is shown in Fig.\ \ref{fig:spectra}. With increasing temperature, the low-temperature modes \textbf{A} and \textbf{B} gradually lose intensity and vanish above $\sim 7$~K, while two new modes, \textbf{C} and \textbf{D},  emerge above $\sim3$~K. This unusual temperature dependence of the ESR spectra is consistent with previous observations~\cite{Zvyagin}. The temperature dependences of the integrated ESR intensities of modes \textbf{A}, \textbf{B}, \textbf{C}, and \textbf{D} (not presented here) reflect that the observed excitations correspond to transitions between excited states (Fig.\ \ref{fig:levels}). It is worth mentioning that the effect of the anisotropy (resulting in the gapped behavior of modes \textbf{A} and \textbf{B} as shown in the inset of Fig.\ \ref{fig:EPR}) is observed only at low temperatures (when thermal fluctuations are suppressed) and disappears with increasing temperature. The frequency-field dependence of the high-temperature modes \textbf{C} and \textbf{D}, measured at 10~K and presented in Fig.\ \ref{fig:EPR10}, can be described by the simple formula $h\nu=g\mu_{B}B$ with $g=2.28$ and $g=2.04$, respectively, where \emph{no} anisotropy is included.  In accordance with exchange-narrowing theory~\cite{Anderson}, for thermally activated states the hopping probability (or exchange frequency) is temperature dependent. At a fixed frequency at low temperatures the concentration of excited triplets is small, while  the hopping probability is high, resulting in a~fast-exchange regime and well-resolved narrow ESR absorptions (modes \textbf{A} and \textbf{B}).  At higher temperatures, the concentration of triplets is increased, the probability of hopping along the ladder legs becomes smaller and the pairs of low-temperature ESR resonances merge to single lines, modes \textbf{C} and \textbf{D} (slow-exchange regime). The crossover from the fast- to slow-exchange limit was observed in a~number of quantum magnets  with spin-singlet ground state, for instance, in TlCuCl$_3$~\cite{Glazkov} and in BaCuSi$_2$O$_6$~\cite{BCSO,Hill}. More details about the exchange narrowing phenomenon in magnetic resonance can be found in Ref.~\onlinecite{Carrington}.

We now discuss the possible origin of anisotropy in BPCB.
Our estimates show that although dipole-dipole interactions can contribute to the anisotropy they are much smaller than the splitting of the modes deduced from our experiments, suggesting  spin-orbit interactions as the main source of the anisotropy. This anisotropy can be caused by antisymmetric Dzyaloshinskii-Moriya (DM) interaction, symmetric anisotropic (sometimes called Kaplan-Shekhtman-Entin-Wohlman-Aharony [KSEWA]) interaction, or by their mixture.  In BPCB, the DM interaction is forbidden on the rungs of the ladder by inversion symmetry, but is allowed along the ladder legs.
The effect of antisymmetric DM and symmetric KSEWA interactions in spin ladders in the weak-coupling limit has been  studied theoretically in Ref.~\onlinecite{anisotropy3}. It was shown that in spin systems with SU(2) symmetry the DM term alone breaks this symmetry, opening a~gap in the excitation spectrum above $B_{c1}$.  On the other hand, the effect of KSEWA interactions is to recover the SU(2) symmetry, leaving the excitation spectrum incommensurate but gapless. No sign of a~gap in the excitation spectrum has been detected by NMR~\cite{NMR_Martin} in the vicinity of $B_{c1}$ and $B_{c2}$ down to 40~mK confirming the applicability of the LL formalism for describing the intermediate phase. The presence of KSEWA interaction on the ladder rungs or a~combination of the DM and KSEWA interactions on the ladder legs (resulting in the anisotropy observed by ESR, but with a~gapless excitation spectrum in the field-induced intermediate phase) explains the experimental results. Our observations call for further development of the theory~\cite{anisotropy3} towards the strong-coupling limit, which is relevant to BPCB.

The effects of the anisotropy, however, need to be taken into account when describing the phase diagram and critical properties of BPCB. For example, $B_{c1}$ and $B_{c2}$ are found to be more sensitive to the direction of the applied magnetic field than expected from the anisotropy of the $g$-tensor only, explaining the variations in the values of the critical fields reported in the literature~\cite{WatsonBPCB,NMR_BPCB,neutrondiffr,BeniBPCB,SaviciINS, Ruegg-Thermo,AnfusoBPCB,Lorenz}. The observation of finite biaxial anisotropy (which breaks the U(1) rotational symmetry) can be of particular importance when applying the magnon  BEC formalism for the description of the field-induced antiferromagnetically ordered phase in BPCB at lower temperatures~\cite{NMR_BPCB,neutrondiffr}. Finally, understanding the role of anisotropy and its experimental consequences in spin ladders itself is of fundamental interest~\cite{anisotropy1, anisotropy2, anisotropy3, anisotropy4, anisotropy5, anisotropy6, anisotropy7}. Therefore, our findings have a~broader impact, offering BPCB as a~model system for investigating also anisotropy  effects in spin ladders.


\emph{Conclusion.---}
The anisotropy of the magnetic interactions in BPCB was unraveled in a~comprehensive ESR study. This anisotropy is an important parameter to be taken into account when describing magnetic properties of this compound. Consequently, BPCB can serve as a~unique model system for the theoretical and experimental investigation of the role of anisotropy (and,  more generally, spin-orbit effects) in spin ladders, in particular close to their quantum critical points.

\emph{Acknowledgments.---}
The authors would like to thank O.~Cepas,  F.~Mila,  and A.I.~Smirnov for  fruitful discussions. Part of this work was supported by the Deutsche Forschungsgemeinschaft,  EuroMagNET (EU contract No. 228043), Swiss National Science Foundation, the Royal Society and the EPRSC. E.\v{C}. is supported by APVV-VVCE-0058-07 and APVV-0006-07.


\begin{thebibliography}{99}
\bibitem{Siegrist} M. Sigrist, T.M. Rice, and F.C. Zhang,   Phys. Rev. B {\bf 49}, 12058  (1994).
\bibitem{scond} E. Dagotto and T.M. Rice, Science  {\bf 271}, 618 (1996).
\bibitem{Uehara} M. Uehara, T. Nagata, J. Akimitsu, H. Takahashi, N. M\^ori, and K. Kinoshita, J. Phys. Soc. Jpn. {\bf 65}, 2764 (1996).
\bibitem{Azuma} M. Azuma, Z. Hiroi, M. Takano, K. Ishida, and Y. Kitaoka, Phys. Rev. Lett. {\bf 73}, 3463 (1994).
\bibitem{Landee} C.P. Landee, M.M. Turnbull, C. Galeriu, J. Giantsidis, and F.M. Woodward, Phys. Rev. B {\bf 63}, 100402(R) (2001).
\bibitem{Masuda}  T. Masuda, A. Zheludev, H. Manaka, L.-P. Regnault, J.-H. Chung, and Y. Qiu, Phys. Rev. Lett. {\bf 96}, 047210 (2006).
\bibitem{WatsonBPCB} B.C. Watson, V.N. Kotov, M.W. Meisel, D.W. Hall, G.E. Granroth, W.T. Montfrooij, S.E. Nagler, D.A. Jensen, R. Backov, M.A. Petruska, G.E. Fanucci, and D.R. Talham,  Phys. Rev. Lett. {\bf 86}, 5168 (2001).
\bibitem{NMR_BPCB} M. Klanj\v{s}ek, H. Mayaffre, C. Berthier, M. Horvati\'{c}, B. Chiari, O. Piovesana, P. Bouillot, C. Kollath, E. Orignac, R. Citro, and T. Giamarchi, Phys. Rev. Lett. {\bf 101}, 137207 (2008).
\bibitem{neutrondiffr} B. Thielemann, Ch. R{\"{u}}egg, K. Kiefer, H.M. R{\o}nnow, B. Normand, P. Bouillot, C. Kollath, E. Orignac, R. Citro, T. Giamarchi, A.M. L{\"{a}}uchli, D. Biner, K. Kr{\"{a}}mer, F. Wolff-Fabris, V. Zapf, M. Jaime, J. Stahn, N.B. Christensen, B. Grenier, D.F. McMorrow, and J. Mesot, Phys. Rev. B {\bf 79}, 020408(R) (2009).
\bibitem{Sologubenko} A.V. Sologubenko, T. Lorenz, J.A. Mydosh, B. Thielemann, H.M. R{\o}nnow, Ch. R{\"{u}}egg, and K.W. Kr{\"{a}}mer, Phys. Rev. B {\bf 80}, 220411(R) (2009).
    \bibitem{Ruegg-Thermo} Ch. R{\"{u}}egg, K. Kiefer, B. Thielemann, D. F. McMorrow, V. Zapf, B. Normand, M.B. Zvonarev, P. Bouillot, C. Kollath, T. Giamarchi, S. Capponi, D. Poilblanc, D. Biner, and K.W. Kr{\"{a}}mer, Phys. Rev. Lett. {\bf 101}, 247202 (2008).
\bibitem{BeniBPCB} B. Thielemann, Ch. R{\"{u}}egg, H.M. R{\o}nnow, A.M. L{\"{a}}uchli, J.-S. Caux, B. Normand, D. Biner, K.W. Kr{\"{a}}mer, H.-U. G{\"{u}}del, J. Stahn, K. Habicht, K. Kiefer, M. Boehm, D.F. McMorrow, and J. Mesot, Phys. Rev. Lett. {\bf 102}, 107204 (2009).
\bibitem{anisotropy1} E. Orignac and T. Giamarchi, Phys. Rev. B {\bf 57}, 5812 (1998).
\bibitem{anisotropy2} A.P. Tonel, A. Foerster, J. Links, and A.L. Malvezzi, Phys. Rev. B {\bf 64}, 054420 (2001).
\bibitem{anisotropy3} R.~Citro and E.~Orignac, Phys. Rev. B {\bf 65}, 134413 (2002),
\bibitem{anisotropy4} Z.J. Ying, I. Roditi, A. Foerster, and B. Chen , Eur. Phys. J. B {\bf 41}, 67 (2004);
\bibitem{anisotropy5} E. Orignac, R. Citro, S. Capponi, and D. Poilblanc, Phys. Rev. B {\bf 76}, 144422 (2007);
\bibitem{anisotropy6} K. Penc, J.-B. Fouet, S. Miyahara, O. Tchernyshyov, and F. Mila, Phys. Rev. Lett. {\bf 99}, 117201 (2007);
\bibitem{anisotropy7} S. Miyahara, J.-B. Fouet, S.R. Manmana, R.M. Noack, H. Mayaffre, I. Sheikin, C. Berthier, and F. Mila, Phys. Rev. B {\bf 75}, 184402 (2007).
\bibitem{LL} F.D.M. Haldane, Phys. Rev. Lett. {\bf 45}, 1358  (1980).
\bibitem{BEC} T. Giamarchi and A.M. Tsvelik, Phys. Rev. B {\bf 59}, 11398 (1999);  T. Nikuni, M. Oshikawa, A. Oosawa, and H. Tanaka, Phys. Rev. Lett. {\bf 84}, 5868 (2000); S. Wessel, M. Olshanii, and S. Haas, Phys. Rev. Lett. {\bf 87}, 206407 (2001).
\bibitem{Patyal} B.R. Patyal, B.L. Scott, and R.D. Willett,  Phys. Rev. B {\bf 41}, 1657 (1990).
\bibitem{SaviciINS} A.T. Savici, G.E. Granroth, C.L. Broholm, D.M. Pajerowski, C.M. Brown, D.R. Talham, M.W. Meisel, K.P. Schmidt, G.S. Uhrig, and S.E. Nagler, Phys. Rev. B {\bf 80}, 094411 (2009).
\bibitem{AnfusoBPCB} F. Anfuso, M. Garst, A. Rosch, O. Heyer, T. Lorenz, Ch. R{\"{u}}egg, and K. Kr{\"{a}}mer, Phys. Rev. B {\bf 77}, 235113 (2008).
\bibitem{Lorenz} T. Lorenz,  O. Heyer, M. Garst, F. Anfuso, A. Rosch, Ch. R{\"{u}}egg, and K. Kr{\"{a}}mer, Phys. Rev. Lett. {\bf 100}, 067208 (2008).
\bibitem{Lemmens} K.-Y. Choi, V. Gnezdilov, B.C. Watson, M.W. Meisel, D.R. Talham, and P. Lemmens, J. Phys.: Condens. Matter {\bf 17}, 4237 (2005).
\bibitem{spectrometer} S.A. Zvyagin, J. Krzystek, P.H.M. van Loosdrecht, G. Dhalenne, and A. Revcolevschi, Physica B {\bf 346-347}, 1 (2004).
\bibitem{noteD} We found the same angular dependences both in the hydrogenated as well as in the deuterated crystals.
\bibitem{Bleaney} B. Bleaney and D.K. Bowers, Proc. Roy. Soc (London) {\bf A214}, 451 (1952).
\bibitem{Abragam} A. Abragam and B. Bleaney, Electron Paramagnetic Resonance of Transition Ions (Dover, New York, 1970).
\bibitem{easyspin} S. Stoll and A. Schweiger, J. Magn. Reson. {\bf 178}, 42 (2006).
\bibitem{Zvyagin} S. Zvyagin, B.C. Watson, J.-H. Park, D.A. Jensen, A. Angerhofer, L.-C. Brunel, D.R. Talham, and M.W. Meisel,  Physica B {\bf 329-333}, 1211 (2003).
\bibitem{intensity} For frequencies above 150~GHz the calculated intensities of the transitions shown by the dotted lines are less than 0.5\% of those for the modes \textbf{A} and \textbf{B}.
\bibitem{Anderson} P.W. Anderson, J. Photogr. Sci. {\bf 9}, 316 (1954).
\bibitem{Glazkov} V.N.~Glazkov, A.I. Smirnov, H. Tanaka, and A. Oosawa, Phys. Rev. B {\bf 69}, 184410 (2004).
\bibitem{BCSO} S.A.~Zvyagin, J. Wosnitza, J. Krzystek, R. Stern, M. Jaime, Y. Sasago, and K. Uchinokura, Phys. Rev. B {\bf 73}, 094446 (2006).
\bibitem{Hill} S.E.~Sebastian, P. Tanedo, P.A. Goddard, S.-C. Lee, A. Wilson, S. Kim, S. Cox, R. D. McDonald, S. Hill, N. Harrison, C.D. Batista, and I.R. Fisher, Phys. Rev. B {\bf 74}, 180401(R) (2006).
\bibitem{Carrington} A.~Carrington and A.D.~McLachlan, $Introduction$ $to$ $Magnetic$ $Resonance$ (Harper and Row, New York, 1967).
\bibitem{NMR_Martin} M. Klanj\v{s}ek \textit{et al.}, unpublished.
\end{thebibliography}
\end{document}